\documentclass{article}
\usepackage{spconf,amsmath,graphicx,hyperref,booktabs, amssymb, enumitem, listings, xcolor}
\usepackage[most]{tcolorbox}
\usepackage{cite}

\makeatletter
\let\oldthebibliography\thebibliography
\renewcommand{\thebibliography}[1]{%
  \oldthebibliography{#1}%
  \setlength{\itemsep}{0pt}%
  \setlength{\parsep}{0pt}%
}
\makeatother

\title{Spatial Transfusion: From Structured Spatial Scene Understanding to Generation and Editing}

\name{Jaeyeon Kim$^{1*}$\thanks{$^*$Work done during an internship at Microsoft}  \quad Sebastian Braun$^2$ \quad Hannes Gamper$^2$}
\address{$^1$Carnegie Mellon University \quad $^2$ Microsoft Corporation}

\begin{document}
\ninept

\maketitle

\begin{abstract}
Spatial audio conveys not only what sounds are present, but where they originate and how they interact with the environment. Recent audio-language models increasingly understand and reason about such scenes, yet spatial audio understanding, generation, and editing remain largely separate. We present Spatial Transfusion, a unified first-order ambisonics (FOA) framework that extends a pretrained spatial audio understanding model to generation and editing through flow matching over VAE latents. We formulate editing as a structured scene transformation over a taxonomy of atomic edits. Keeping the pretrained backbone frozen, we introduce modality-based LoRA routing and a curriculum that progressively adds generation, editing, and natural-language interaction. Experiments show that Spatial Transfusion largely preserves spatial understanding while outperforming full fine-tuning in generation and editing. The resulting model supports controllable FOA generation and editing through both structured and natural-language interfaces. Audio samples are available at \url{https://microsoft.github.io/spatial-transfusion/}.

\end{abstract}
\begin{keywords}
Audio-Language Model, Unified Framework, Transfusion, Spatial Audio Editing, First-Order Ambisonics
\end{keywords}

\section{Introduction}

Humans perceive the acoustic world spatially. Spatial audio provides an immersive representation of acoustic scenes by capturing not only what sounds are present, but also where they originate and how they interact with the surrounding environment. Compared with conventional mono audio, spatial audio scenes involve additional factors such as source locations and distances, room acoustics, and diffuse background components, which make both understanding and generation substantially more challenging. Recent large audio-language models (LALMs) have started extending their capabilities to spatial audio by integrating a spatial audio encoder into a large language model (LLM), enabling description of and reasoning about complex acoustic environments from binaural~\cite{bat, owls}, first-order ambisonics (FOA)~\cite{tang2024can, sciphi, spatial_omni}, or arbitrary microphone-array~\cite{phasecoder} input.

In parallel, several works have explored unified models that support both understanding and generation in the audio domain, modeling audio either as discrete codec tokens~\cite{ualm,bagpiper} or as continuous latents~\cite{audiochat,uat}. These models share the intuition that understanding and generation are two interfaces to a shared underlying representation, so that generative capability can be built on top of representations learned for perception~\cite{ualm,bagel}.

In spatial audio, however, understanding, generation, and editing are mainly addressed by specialized models rather than a unified framework. Existing spatial audio generation methods are conditioned on text~\cite{immersediffusion,sonicmotion,bewo} or video~\cite{visage, visage_sc}, but offer limited per-source control in multi-source scenes, where each source has its own direction, distance, and temporal activity. Editing has received even less attention: recent methods have begun to target stereo~\cite{smartdj,audiochat}, but their spatial operations are limited to left-right panning without control over full 3D direction or separate spatial treatment of directional sources and diffuse background.
This motivates a natural question: can the spatial scene understanding of an LALM be extended into a unified framework for generation and editing? We hypothesize that controllable generation and editing benefit from a model that first builds a structured and fine-grained understanding of the acoustic scene. If a model can understand scene attributes such as source positions and temporal activities, the same representations can provide a foundation for controllable synthesis and manipulation.

To investigate this hypothesis, we introduce Spatial Transfusion, the first unified framework that supports understanding, generation, and editing of FOA scenes. We adopt FOA for its playback-agnostic representation and its wide use in both research and consumer platforms. We build Spatial Transfusion on a variant of Sci-Phi~\cite{sciphi}, a recent LALM that produces structured scene descriptions, and extend it to generation and editing under the Transfusion paradigm~\cite{transfusion,bagel,audiochat}. Specifically, we combine an autoregressive understanding objective on text with a flow-matching~\cite{flow_matching,rectified_flow} generation objective on continuous audio latents within a shared Transformer backbone. To support editing, we formulate it as a structured scene transformation and define a taxonomy of atomic edit operations spanning semantic, acoustic, spatial, and temporal dimensions.

A key challenge is that Transfusion-style approaches typically rely on large-scale pretraining with full-model training~\cite{transfusion, bagel, audiochat}, a regime that is impractical for spatial audio, where paired data is far more limited and complex to simulate. Instead, we adapt an existing spatial audio LALM using modality-based LoRA routing, together with a curriculum that progressively extends understanding capabilities to generation and editing. Experiments show that the resulting model largely retains the spatial understanding of the base model while gaining generation and editing through structured interfaces, and that instruction tuning further extends these capabilities to natural-language interaction. Notably, our adaptation substantially outperforms full fine-tuning in generation and editing, and an ablation shows that routing by modality is the most balanced choice among adaptation strategies.
\section{Spatial Transfusion}
\begin{figure*}[t]
\centering
\includegraphics[width=\textwidth]{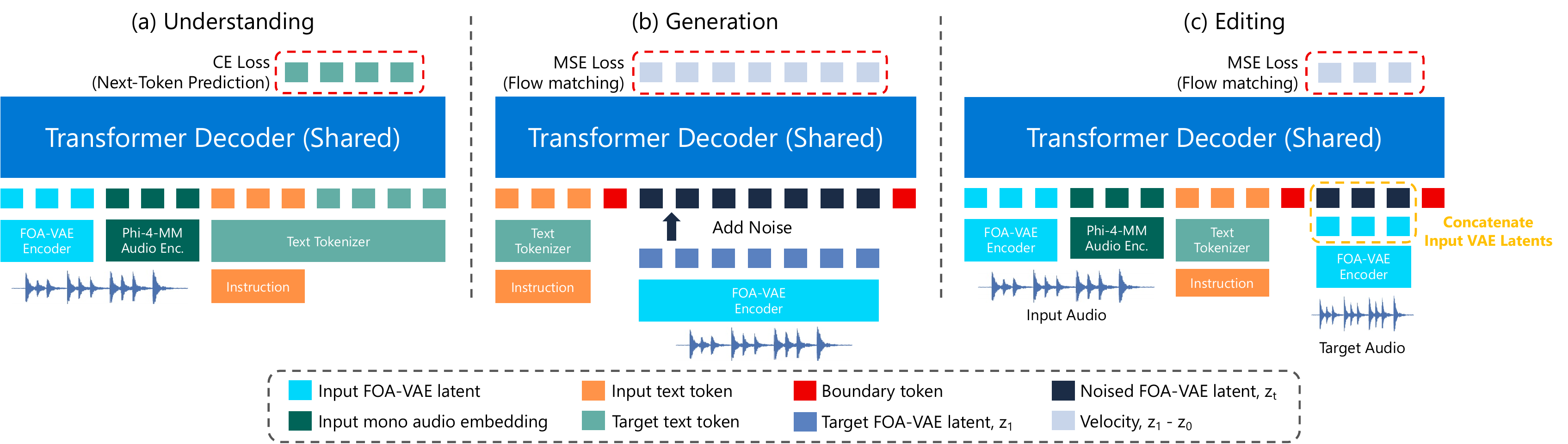}
\vspace{-1.5\baselineskip}
\caption{Overview of Spatial Transfusion. A shared Transformer backbone serves (a) understanding with a cross-entropy loss on text, and (b) generation and (c) editing with a flow-matching loss on target FOA-VAE latents. In editing, clean source latents are additionally concatenated channel-wise with the noised target latents.}
\label{fig:arch}
\end{figure*} 

\subsection{Background} 
\label{sec:background}

\noindent\textbf{FOA-VAE.} To model FOA audio in a compact latent space, we adopt FOA-VAE~\cite{foavae}, a continuous variational autoencoder for four-channel FOA at 16~kHz. It operates on the complex STFT, encoding the real and imaginary parts of all channels into a continuous latent with a fully convolutional encoder and reconstructing the spectrum with a symmetric decoder. Beyond the standard reconstruction, adversarial, and KL objectives, it is trained with a spatial consistency loss that penalizes angular mismatch between the intensity vectors~\cite{dirac} of the input and the reconstruction, so that directional cues survive compression. The resulting latent runs at 7.8~Hz with 512 dimensions each for the mean and variance, retaining both semantic content and perceptually meaningful spatial cues.

\noindent\textbf{Sci-Phi.} Sci-Phi~\cite{sciphi} extends Phi-4-Multimodal (Phi-4-MM)~\cite{[phi4mm]}, which supports mono audio understanding, to spatial acoustic scenes by adding a spatial audio encoder and training an additional LoRA within the LLM. Given an FOA recording, it produces a structured scene description that enumerates the room, a diffuse background, and up to five directional sources with their semantic, acoustic, spatial, and temporal attributes, as shown in Fig.~\ref{fig:json}. 
Training data is simulated by convolving sound events with room impulse responses from image-source rooms~\cite{allen1979image}, at random locations, levels, and onsets, and mixing them with a diffuse background; a structured scene description lists these parameters in JSON text format.

Our work starts from a variant of Sci-Phi that replaces the original spatial encoder with the FOA-VAE encoder. The latent mean and variance are concatenated channel-wise, projected to the LLM embedding dimension, and modeled jointly with the mono audio and text embeddings through a LoRA of rank 16. This understanding path is architecturally retained in Spatial Transfusion (Fig.~\ref{fig:arch}(a)).

\subsection{Architecture}

\noindent\textbf{Transfusion-Based Modeling.} Spatial Transfusion extends Sci-Phi into a unified model that supports understanding, generation, and editing within a shared Transformer backbone (Fig.~\ref{fig:arch}). Following the Transfusion paradigm~\cite{transfusion, bagel}, text tokens are modeled autoregressively with a cross-entropy objective under causal attention, while target FOA-VAE latents are modeled with a flow-matching objective under bidirectional attention, since they are denoised jointly rather than emitted one at a time. Specifically, the target audio is encoded by the FOA-VAE into a posterior $\mathcal{N}(\mu, \sigma^2)$, from which we sample a 512-dimensional target latent $z_1$. Given noise $z_0 \sim \mathcal{N}(0, I)$ and $t$ drawn from a logit-normal distribution~\cite{sd3}, we form $z_t = (1-t) z_0 + t z_1$ and train the model to predict the velocity~\cite{flow_matching, rectified_flow}:
\begin{equation}
\mathcal{L}_{\mathrm{FM}} = \mathbb{E}_{t, z_0, z_1} \left\| v_\theta(z_t, t, c) - (z_1 - z_0) \right\|^2,
\end{equation}
where $c$ is the conditioning context. The two objectives are combined as $\mathcal{L} = \lambda \mathcal{L}_{\mathrm{CE}} + \mathcal{L}_{\mathrm{FM}}$.

The same backbone serves all three tasks, which differ only in what occupies the context. Understanding takes an FOA input and an optional text instruction, and decodes either a scene description or an answer. Generation takes a scene description and predicts the velocity toward the target latents. Editing takes a source scene together with an edit instruction, and predicts the velocity toward the latents of the edited scene. In editing, the source scene is conditioned through two paths: it is passed through the understanding path as a conditioning input, and, since this alone fails to preserve the input audio content, the clean source latents are additionally concatenated channel-wise with the noised target latents, following common practice in image and audio editing~\cite{ip2p, smartdj}. Three modules are added for the latent path: a linear projector that maps noised latents (and, in editing, the channel-concatenated source latents) into the LLM embedding space, a timestep embedding that conditions on $t$, and a linear head that predicts the velocity from the final hidden states. All three are randomly initialized and add about 13M parameters.

All three tasks are trained jointly on mixed batches, with the cross-entropy loss applied to output text tokens and the flow-matching loss to target latents. We set $\lambda = 0.1$ to balance the two capabilities. At inference, the task is determined by the context: the model either decodes the output text autoregressively, or samples the target latents by integrating the predicted velocity field from $z_0$ to $z_1$, which the FOA-VAE decoder converts back to a waveform.

\label{sec:routing}
\noindent\textbf{Modality-Based LoRA Routing.} Rather than fully fine-tuning the Sci-Phi backbone, we adapt it with LoRA~\cite{lora}. Instead of applying a single LoRA to all tokens, we route tokens to different LoRA adapters, similar to the Mixture-of-Transformers (MoT) structure~\cite{mot} used by recent Transfusion models~\cite{lmfusion, bagel}. In MoT, tokens are assigned to paths by their type, where each path has separate weights for the query, key, and value projections, feed-forward layers, and normalization. These paths interact only through the shared self-attention over the interleaved sequence, so each path has its own parameters while the model still attends across paths.

Whereas MoT replicates the entire backbone per path, we keep the backbone shared and attach a separate LoRA adapter to each path. This gives each path its own parameter space through its adapter without duplicating the backbone.
Among possible routing strategies, we adopt modality-based routing~\cite{mot, lmfusion}: audio tokens, both the input audio embeddings and the target latents, share one path, and text tokens use another. This allows understanding and generation to share audio-adapter parameters, encouraging transfer between the two capabilities. The audio path uses a larger LoRA rank (640 vs.\ 16), since it needs to learn a new generative capability. QK-normalization~\cite{qknorm} is applied in all attention layers for training stability. The new LoRA modules are applied to the LLM weights and trained together with the three new modules. The Phi-4-MM backbone, the encoders, and the existing LoRAs of Sci-Phi remain frozen. In total, the audio adapter adds about 923M parameters and the text adapter 7M, for 943M trainable parameters including the new modules, compared with 3.85B for full fine-tuning.

\subsection{Spatial Audio Editing}

\begin{table}[t]
\centering
\caption{Taxonomy of atomic spatial audio edits by dimension and target. \textit{Move} relocates a source in azimuth, elevation, and distance by selecting a new room impulse response, and thus covers the full 3D position in a single operation.}
\label{tab:taxonomy}
\small
\resizebox{0.8\columnwidth}{!}{\begin{tabular}{lll}
\toprule
\textbf{Dimension} & \textbf{Directional source} & \textbf{Diffuse background} \\
\midrule
Semantic & Add, Remove, Replace           & Add, Remove, Replace \\
Spatial  & Move                           & --- \\
Acoustic & Change level, Pitch shift      & Change level \\
Temporal & Shift time, Time stretch       & --- \\
\bottomrule
\end{tabular}}
\end{table}
\begin{figure}[t]
\centering
\lstset{
  basicstyle=\fontsize{6}{7}\selectfont\ttfamily,
  frame=none, xleftmargin=0pt, xrightmargin=0pt,
  aboveskip=0pt, belowskip=0pt, breaklines=false
}
\tcbset{jsonbox/.style={
  enhanced, colback=gray!5, colframe=black!85, boxrule=0.9pt, arc=3pt,
  left=1.5pt, right=1.5pt, top=3pt, bottom=3pt,
  equal height group=json, valign=center,
  fonttitle=\scriptsize\bfseries, coltitle=white, colbacktitle=black!85,
  attach boxed title to top left={yshift=-2.5mm, xshift=2mm},
  boxed title style={arc=2pt, boxrule=0pt, left=3pt, right=3pt, top=1pt, bottom=1pt},
  toptitle=1pt, before upper=\vspace{1.5mm}
}}
\newsavebox{\jsonfigbox}
\begin{lrbox}{\jsonfigbox}
\begin{minipage}{\columnwidth}
\begin{tcolorbox}[jsonbox, width=0.53\columnwidth, nobeforeafter,
  title={(a) Scene description}]
\begin{lstlisting}
{
 "source_count": 2,
 "RT60": 0.6,
 "room_volume_cubic_meters": 600,
 "background_type": "wind",
 "background_loudness_dBA": -55,
 "sources": [
  {
   "sound_type": "piano",
   "description": "A piano melody
      played softly with chords",
   "start_sec": 6.1,
   "end_sec": 9.3,
   "azimuth": 95,
   "elevation": 8,
   "distance_meters": 2.8,
   "loudness_dBA": -24,
   "C50_dB": 18
  },
  { "sound_type": "bell", ... }
 ]
}
\end{lstlisting}
\end{tcolorbox}\hfill
\begin{tcolorbox}[jsonbox, width=0.455\columnwidth, nobeforeafter,
  title={(b) Edit instruction}]
\begin{lstlisting}
{
 "operation": "move_source",
 "subject": {
  "kind": "source",
  "sound_type": "piano",
  "description": "A piano
    melody played softly..."
 },
 "set": {
  "azimuth": -109,
  "elevation": -20,
  "distance_meters": 0.9
 }
}
\end{lstlisting}
\end{tcolorbox}
\end{minipage}
\end{lrbox}
\resizebox{0.9\columnwidth}{!}{\usebox{\jsonfigbox}}
\vspace{-4pt}
\caption{Structured representations used in Stages 1--3. (a) A scene description lists directional sources with their attributes plus room and background fields; it is the understanding output and the generation input. (b) An edit instruction names an operation from Table~\ref{tab:taxonomy}, its target, and the modification to apply; it is the editing input.}
\label{fig:json}
\end{figure}

\noindent\textbf{Editing Taxonomy.} While generation data can be obtained by simply reversing the understanding examples of Sci-Phi, editing data requires the additional constraint that the input scene must be preserved while only the targeted aspect changes. We therefore formulate spatial audio editing as a structured scene transformation. Extending the atomic-edit formulation of SmartDJ~\cite{smartdj} to the FOA setting, we define a taxonomy of atomic edits spanning semantic, acoustic, spatial, and temporal dimensions, applied to both foreground sources and the diffuse background, resulting in 12 operations. The complete taxonomy is given in Table~\ref{tab:taxonomy}.

\noindent\textbf{Data Simulation.} To generate paired editing data at scale, we extend the Sci-Phi simulation pipeline. Each example is created by perturbing exactly the scene factor targeted by the edit while keeping everything else fixed: moving a source replaces its room impulse response, replacing a source swaps its audio file, and so on. Re-rendering the perturbed scene yields an (input, instruction, target) triplet with exact ground truth for both the edited and the preserved content, which we generate for every operation in the taxonomy.

\subsection{Training Curriculum}
\label{sec:curriculum}
We adopt a four-stage curriculum that extends one capability at a time for effective adaptation:
\begin{itemize}[leftmargin=*, itemsep=0.3pt, topsep=1pt]
  \item \textbf{Stage 1: Single-source understanding
  and generation.} The model learns to map single-source FOA scenes to structured descriptions and back, acquiring basic generative capability.
  \item \textbf{Stage 2: Single-source editing.} Editing is added next, since an edit is a conditional generation that must also preserve the input.
  \item \textbf{Stage 3: Multi-source scenes.} All three tasks are extended to scenes with up to five directional sources.
  \item \textbf{Stage 4: Instruction tuning.} The JSON-based capabilities are mapped onto free-form natural language: for each of understanding, generation, and editing, half of the samples use JSON and half use natural-language captions or instructions. Question-answering (QA) samples are additionally added to understanding.
\end{itemize}

The first three stages use structured JSON (Fig.~\ref{fig:json}) rather than free-form text. JSON defines every attribute explicitly, which gives fine-grained control for generation, and matches the format Sci-Phi already produces. Moreover, mapping JSON to captions is easier than the reverse, as captions may lack explicit scene details. Learning the structured form first thus supports both interfaces with different levels of controllability.

Table~\ref{tab:curriculum} lists the data at each stage. Stage 3 keeps 1.0M of the Stage 1--2 samples and adds 2.6M multi-source scenes; for editing, it keeps 0.45M of the Stage 2 samples and adds 0.95M. Stage 4 draws half of its samples from Stage 3, with natural-language captions generated by GPT-5.4-mini~\cite{gpt54mini}, and adds QA pairs from Sci-Phi-Chat~\cite{sciphichat}. In total, 4.6M unique scenes are used for understanding and generation and 2.4M for editing.

\begin{table}[t]
\centering
\caption{Training curriculum. Sampling ratio is given as understanding\,:\,generation\,:\,editing\,:\,QA. Sample counts are the number of unique pairs available at each stage.}
\label{tab:curriculum}
\small
\setlength{\tabcolsep}{5pt}
\resizebox{0.75\columnwidth}{!}{
\begin{tabular}{ccccc@{\hspace{10pt}}l}
\toprule
\textbf{Stage} & \textbf{Steps} & \textbf{Und/Gen} & \textbf{Edit} & \textbf{QA} & \textbf{Ratio} \\
\midrule
1 & 30K  & 2.0M & --   & --   & 0.2\,:\,0.8\,:\,0\,:\,0 \\
2 & 70K  & 2.0M & 1.5M & --   & 0.2\,:\,0.3\,:\,0.5\,:\,0 \\
3 & 100K & 3.6M & 1.4M & --   & 0.2\,:\,0.4\,:\,0.4\,:\,0 \\
4 & 50K  & 1.8M & 0.7M & 250K & 0.2\,:\,0.3\,:\,0.3\,:\,0.2 \\
\bottomrule
\end{tabular}}
\end{table}
\section{Experiments}
\label{sec:experiments}
\begin{table}[t]
\centering
\caption{Understanding results on the FOA-MEIR-based test set.}
\label{tab:understanding_results}
\footnotesize
\setlength{\tabcolsep}{2pt}
\resizebox{\columnwidth}{!}{%
\begin{tabular}{lccccc}
\toprule
\textbf{Model} & \textbf{Tuple}$\uparrow$ & \textbf{Nsrc Acc.}$\uparrow$ & \textbf{Src CLAP}$\uparrow$ & \textbf{Dir. Err.}$\downarrow$ & \textbf{Time IoU}$\uparrow$ \\
\midrule
Sci-Phi                & 0.798 & \textbf{0.674} & 0.670 & 13.6 & \textbf{0.754} \\
\midrule
Full fine-tuning       & 0.805 & 0.654 & 0.679 & \textbf{12.6} & 0.726 \\
Spatial Transfusion   & \textbf{0.812} & 0.635 & \textbf{0.683} & 14.3 & \textbf{0.754} \\
\bottomrule
\end{tabular}%
}
\end{table}

\subsection{Experimental Setup}
\label{sec:experimental_setup}

\noindent\textbf{Configuration.} We train with AdamW at a learning rate of $1\times10^{-4}$, with linear warmup followed by cosine decay to $1\times10^{-5}$. Optimization settings are held constant across curriculum stages. At inference, understanding uses greedy decoding, while generation and editing use 50 flow-matching steps with classifier-free guidance~\cite{cfg} applied to the text condition at scales of 3.5 and 4.0, respectively.

\noindent\textbf{Baselines.} Evaluation uses a held-out set rendered with sources, real room impulse responses, and ambient noise from the FOA-MEIR dataset~\cite{meir}. Existing spatial audio generators differ in their conditioning interfaces and do not support our structured scene specification.
To assess the extension from understanding to generation and editing, we instead compare Spatial Transfusion with full fine-tuning of the same pretrained backbone, using the same training data and curriculum and a learning rate of $5\times10^{-5}$.
This provides a controlled comparison of adaptation strategies. 

The Sci-Phi variant (Sec.~\ref{sec:background}) serves as the understanding baseline, and FOA-VAE reconstruction is the generation topline. For editing, we include copy-input and target-JSON baselines to contrast input preservation with scene regeneration. Copy-input returns the source audio unchanged, whereas target-JSON uses Spatial Transfusion to generate audio solely from the ground-truth target scene description. Ground-truth target audio serves as the editing reference. Unless noted, we evaluate the model after Stage 3, with structured JSON as the understanding output and generation and editing input.

\noindent\textbf{Metrics.} Understanding follows the Sci-Phi protocol~\cite{sciphi}: Tuple Score summarizes semantic, spatial, and temporal accuracy, while source-count accuracy (Nsrc Acc.), source-level CLAP similarity (Src CLAP), direction error (Dir.\ Err.), and temporal IoU (Time IoU) provide detailed assessments. Generation is evaluated using FAD-PSELD for spatial--semantic fidelity, FAD-CLAP for audio fidelity, and CLAP-TA for alignment with the target description. FOA covariance loss~\cite{cov_loss} assesses spatial-channel structure via the L1 distance between the normalized channel covariance matrices of generated and target audio. We use FAD-PSELD, computed with PSELDNets~\cite{pseldnets}, as the primary spatial metric following a recent sensitivity analysis~\cite{kamath2026sensitivity}. Editing additionally uses EditFLAM~\cite{audiochat}, based on operation-specific changes in FLAM~\cite{flam} probabilities between input and edited audio, and CLAP-AA for similarity between these signals. For editing, EditFLAM and CLAP-TA are evaluated only on operations that change source or background semantic content. All CLAP-based metrics use MS-CLAP~\cite{msclap}, and FAD-CLAP is computed with fadtk~\cite{fadtk}.

\subsection{Results}
\label{sec:results}

\begin{table}[t]
\centering
\caption{Generation results on the FOA-MEIR-based test set.}
\label{tab:generation_results}
\setlength{\tabcolsep}{3pt}
\resizebox{\columnwidth}{!}{%
\begin{tabular}{lcccc}
\toprule
\textbf{Model} & \textbf{FAD-PSELD}$\downarrow$ & \textbf{FAD-CLAP}$\downarrow$ & \textbf{CLAP-TA}$\uparrow$ & \textbf{FOA Cov.}$\downarrow$ \\
\midrule
Full fine-tuning       & 3.82 & 379 & 0.232 & 0.196 \\
Spatial Transfusion   & \textbf{0.964} & \textbf{229} & \textbf{0.264} & \textbf{0.166} \\
\midrule
VAE recon.\ (topline) & 0.103 & 47.6 & 0.340 & 0.013 \\
\bottomrule
\end{tabular}%
\vspace{-3mm}   
}
\end{table}

\begin{table}[t]
\centering
\caption{Editing results on the FOA-MEIR-based test set.}
\label{tab:editing_results}
\setlength{\tabcolsep}{3pt}
\resizebox{\columnwidth}{!}{%
\begin{tabular}{lcccc}
\toprule
\textbf{Model} & \textbf{EditFLAM}$\uparrow$ & \textbf{FAD-PSELD}$\downarrow$ & \textbf{CLAP-AA}$\uparrow$ & \textbf{CLAP-TA}$\uparrow$ \\
\midrule
Copy-input             & 0.0 & 0.005 & 1.000 & 0.324 \\
Target-JSON gen. & 19.7 & 1.19 & 0.701 & 0.248 \\
\midrule
Full fine-tuning       & 15.5 & 0.596 & 0.795 & 0.275 \\
Spatial Transfusion   & \textbf{17.8} & \textbf{0.323} & \textbf{0.841} & \textbf{0.306} \\
\midrule
Ground truth (topline) & 28.9 & 0.000 & 0.913 & 0.344 \\
\bottomrule
\end{tabular}%
}
\end{table}

\noindent\textbf{Main results.}
Table~\ref{tab:understanding_results} shows that Spatial Transfusion largely retains the original Sci-Phi model's understanding performance. Full fine-tuning performs similarly. These results indicate that adding generation and editing within the unified framework does not compromise spatial understanding. For generation (Table~\ref{tab:generation_results}), Spatial Transfusion outperforms full fine-tuning on all metrics, reducing FAD-PSELD from 3.82 to 0.964. Although a gap remains to FOA-VAE reconstruction, these results favor adaptation with modality-based LoRA routing over full fine-tuning in our setup.

For editing (Table~\ref{tab:editing_results}), target-JSON generation uses the ground-truth target scene description and achieves higher EditFLAM than Spatial Transfusion, but its lower CLAP-AA and higher FAD-PSELD indicate poorer input preservation and distributional fidelity. Copy-input attains near-zero FAD-PSELD because atomic edits leave most of the scene unchanged, yet its zero EditFLAM reflects the absence of edits. Spatial Transfusion achieves positive EditFLAM while maintaining higher CLAP-AA and lower FAD-PSELD than target-JSON generation, indicating that it applies edits while better preserving the input scene. It also outperforms full fine-tuning on all editing metrics, supporting our adaptation.

\begin{table}[t]
\centering
\caption{Generalization to natural-language interfaces. Stage 4 is evaluated with both JSON and natural-language (NL) inputs. Sci-Phi\,+\,GPT answers questions by prompting GPT-5.4-mini with Sci-Phi's description. QA is evaluated on the MASU benchmark from Sci-Phi-Chat.}
\label{tab:nl_results}
\setlength{\tabcolsep}{3pt}
\resizebox{\columnwidth}{!}{%
\begin{tabular}{lccccc}
\toprule
& \textbf{QA} & \textbf{Und.} & \textbf{Gen.} & \multicolumn{2}{c}{\textbf{Edit}} \\
\cmidrule(lr){2-2}\cmidrule(lr){3-3}\cmidrule(lr){4-4}\cmidrule(lr){5-6}
\textbf{Model} & \textbf{Acc.}$\uparrow$ & \textbf{Tuple}$\uparrow$ & \textbf{FAD-PSELD}$\downarrow$ & \textbf{EditFLAM}$\uparrow$ & \textbf{FAD-PSELD}$\downarrow$ \\
\midrule
Sci-Phi\,+\,GPT      & 0.208 & --    & --   & --    & --    \\
\midrule
Stage 3 (JSON)       & --    & 0.812 & 0.964 & 17.8 & 0.323 \\
Stage 4 (JSON)       & --    & 0.813 & 0.991 & 18.7 & 0.620 \\
Stage 4 (NL)         & 0.476 & --    & 1.39 & 20.9 & 0.772 \\
\bottomrule
\end{tabular}%
\vspace{-3mm}
}
\end{table}

\noindent\textbf{Natural-language generalization.}
Table~\ref{tab:nl_results} shows that Stage 4 instruction tuning largely retains the model's structured capabilities, with some loss in editing fidelity. With natural-language inputs, generation and editing remain broadly comparable to their JSON counterparts, although fidelity decreases, possibly due to less explicit scene specifications. On QA, the instruction-tuned model outperforms the Sci-Phi\,+\,GPT cascade, demonstrating that the adapted model can be extended to spatial question answering. Together, these results show that our training curriculum extends structured scene capabilities to free-form natural-language interaction while largely preserving performance with the structured interface.

\begin{table}[t]
\centering
\caption{Ablation on LoRA routing. \textbf{Best}; \underline{second best}.}
\label{tab:ablation_results}
\setlength{\tabcolsep}{3pt}
\resizebox{\columnwidth}{!}{%
\begin{tabular}{lcccc}
\toprule
& \textbf{Und.} & \textbf{Gen.} & \multicolumn{2}{c}{\textbf{Edit}} \\
\cmidrule(lr){2-2}\cmidrule(lr){3-3}\cmidrule(lr){4-5}
\textbf{Adaptation} & \textbf{Tuple}$\uparrow$ & \textbf{FAD-PSELD}$\downarrow$ & \textbf{EditFLAM}$\uparrow$ & \textbf{FAD-PSELD}$\downarrow$ \\
\midrule
Dense LoRA             & 0.798 & \underline{1.13} & \textbf{18.8} & 1.13 \\
Task-based routing     & \textbf{0.814} & 1.40 & 9.8 & \textbf{0.321} \\
Modality-based routing & \underline{0.812} & \textbf{0.964} & \underline{17.8} & \underline{0.323} \\
\bottomrule
\end{tabular}%
}
\end{table}

\noindent\textbf{Ablation on LoRA routing.}
Table~\ref{tab:ablation_results} compares modality-based routing with two alternative adaptation strategies. \textit{Dense LoRA} applies a single rank-640 LoRA to all tokens, giving 936M trainable parameters. \textit{Task-based routing} follows the understanding--generation split used in BAGEL~\cite{bagel}: input audio embeddings and text use the understanding path, while target latents use the generation path. These alternatives exhibit different trade-offs: Dense LoRA favors edit effectiveness at the expense of scene fidelity, whereas task-based routing maintains understanding but yields weaker generation and less effective editing. Modality-based routing provides the most balanced performance across tasks, motivating its use in Spatial Transfusion.

\section{Conclusion}
We presented Spatial Transfusion, a unified FOA framework that extends spatial audio understanding to generation and editing. With the pretrained backbone frozen, modality-based LoRA routing outperforms full fine-tuning in generation and editing while preserving understanding. Our curriculum extends these capabilities to natural language, supporting both structured and free-form interfaces. Future work includes extension to real-world recordings.

\section{Compliance with Ethical Standards}
This study uses simulated scenes and publicly available datasets. No ethical approval was required.

\section{Acknowledgments}
This work was conducted during an internship at Microsoft. S.~Braun and H.~Gamper are employees of Microsoft Corporation. The authors have no other relevant financial or non-financial interests to disclose. AI tools were used for language editing.

\bibliographystyle{IEEEbib}
\bibliography{main}

\end{document}